\documentclass[aps,twocolumn,showpacs]{revtex4-1}
\usepackage{graphicx}
\usepackage{amsmath,amsfonts}
\usepackage[T1]{fontenc}

\begin{document}

\title{Dynamic structure factor and drag force in a one-dimensional strongly-interacting Bose gas at finite temperature}

\author{Guillaume Lang}
\affiliation{Universit\'e Grenoble Alpes, LPMMC, F-38000 Grenoble, France}
\affiliation{CNRS, LPMMC, F-38000 Grenoble, France}
\author{Frank Hekking}
\affiliation{Universit\'e Grenoble Alpes, LPMMC, F-38000 Grenoble, France}
\affiliation{CNRS, LPMMC, F-38000 Grenoble, France}
\author{Anna Minguzzi}
\affiliation{Universit\'e Grenoble Alpes, LPMMC, F-38000 Grenoble, France}
\affiliation{CNRS, LPMMC, F-38000 Grenoble, France}

\begin{abstract}
 We study the effect of thermal and quantum fluctuations on the dynamical response of a one-dimensional strongly-interacting Bose gas in a tight atomic waveguide. We combine the Luttinger liquid theory at arbitrary interactions and the exact Bose-Fermi mapping in the Tonks-Girardeau-impenetrable-boson limit to obtain the dynamic structure factor of the strongly-interacting fluid at finite temperature. Then, we determine the drag force felt by a potential barrier moving along the fluid in the experimentally realistic situation of finite barrier width and temperature.
 
 \end{abstract}

\maketitle

\section{Introduction}

Superfluidity is one of the most dramatic manifestations of quantum many-body physics. The question whether a degenerate neutral bosonic fluid should be considered as superfluid or not is far from trivial, especially in one dimension, and it is agreed that superfluidity is rather a collection of phenomena than a well-defined phenomenon. In the last decades, one-dimensional (1D) systems, in which the effect of interactions is known to be enhanced compared to 3D, have attracted an increasing interest \cite{Bloch,Zwerger,Naegerl,Bouchoule,Van Druten,Paredes}. In particular, cold atomic gases can now be confined to investigate the effects of low dimensionality \cite{Paredes,Roati,Billy,Esslinger,Chen}. An issue is the effect of the interactions in the degenerate fluid on the possibility of a superfluid flow. While several criteria lead to the same conclusions in 3D \cite{Leggett,Leggett2,Nozieres,Stringari,Hess,Abo,Hodby}, various aspects of coherence and superfluidity may be characterized in one-dimensional systems by different observables. For instance, there is no Bose-Einstein condensation in 1D owing to the Mermin-Wagner theorem, yet there may still subsist off-diagonal quasi-long-range order \cite{Hohenberg, Haldane}.

Among other possibilities, superfluidity can be characterized by the absence of a (viscous) drag force acting on a moving fluid when it encounters an impurity or a potential barrier. The case of a mobile impurity has been the object of intense activity in the last years \cite{Gangardt, Bonart, Minardi, Fukuhara, Kantian, Arzamasovs, Fazio}. Here, we focus on an external potential barrier driven at constant velocity across the fluid. In this configuration, the drag force concept has already been used to probe superfluidity in a 2D Bose gas \cite{Dalibard}. 
With an idealized model of a delta-potential, Pitaevskii and Astrakharchik \cite{Pitaevskii} showed that, according to the drag force criterion, quantum fluctuations give rise to a breakdown of superfluidity at large interactions in a 1D Bose gas, while the fluid may well exhibit a behavior close to superfluid when interactions are small. The calculation of the drag force in linear response theory requires the knowledge of the dynamic structure factor. This itself is an important measurable quantity which shows the many-body spectrum of collective excitations in the fluid, and is experimentally accessible by Bragg scattering \cite{Stenger,Ozeri,Calabrese,David,Brunello,Esslinger,Zambelli}.

In this work we study the dynamic structure factor and the drag force under the experimentally relevant conditions of finite temperature and finite width of the barrier.

We describe the system by the Lieb-Liniger model of interacting 1D bosons \cite{Lieb}, now realizable with experiments with ultracold atomic gases in a wide range of interaction strength \cite{Kinoshita1,Citro}. We focus on the strongly interacting regime and describe it using two complementary techniques: the Tomonaga-Luttinger liquid (LL) and the exact Tonks-Girardeau solution (TG). The LL approach is valid at low energies and temperatures, for intermediate to strong interactions. The TG exact solution describes impenetrable bosons with infinite repulsive interactions by means of a Bose-Fermi mapping.

First, we determine the dynamic structure factor using the LL and TG methods, thus obtaining its temperature dependence for an interaction regime where the Bogoliubov approximation \cite{Sykes} is not applicable. Our approach is particularly suitable to describe the umklapp region of momenta $q$ in the vicinity of 2$k_F$, where $k_F$ is the Fermi wavevector of the mapped Fermi gas. In this region, a Random Phase Approximation perturbative approach \cite{Cheon,Cherny,ChernyCaux,ChernyBrand} is not applicable and predicts a non regular behavior for the dynamic structure factor. Although the dynamic structure factor is amenable to Bethe Ansatz calculations \cite{Panfil,ChernyBrandbis}, to date the strongly interacting regime has not been covered by this technique at finite temperature. In our work, specifically we focus on the comparison between the results obtained with the LL and TG approaches. In order to achieve this goal we determine the parameters of the LL model by solving the Bethe Ansatz equations for the Lieb-Liniger model both at zero and finite temperature. Then, in a linear response approach, we determine the drag force behavior as a function of the barrier velocity, both at zero and finite temperature, including the effect of a finite width of the barrier.

The paper is organized as follows: in Section \ref{Concepts} we present the system and the concepts of dynamic structure factor and drag force; in Sections \ref{TG} and \ref{LL} we evaluate them for a Tonks-Girardeau gas and a Luttinger liquid respectively.

\section{The system, and concepts to probe its superfluidity}
\label{Concepts}
We consider a system of $N_0$ ultracold spinless interacting bosonic atoms confined in a one-dimensional waveguide, as described by the Hamiltonian
\begin{equation}
\label{LiebLiniger}
\small{\!\!\!H_0=\int_0^L\!\!\mathrm{d} x\left[\frac{-\hbar^2}{2m}\psi^{\dagger} \partial_x^2\psi +\frac{1}{2}\int_0^L\!\!\mathrm{d}x'V(x-x')n(x)n(x')\right]}\!,
\end{equation}
where $L$ is the length of the system, $m$ the mass of an atom, $\psi$ a bosonic field operator satisfying the commutation relation $[\psi(x),\psi^\dagger(x')]=\delta(x-x')$, and $n(x)=\psi^\dagger(x)\psi(x)$ the density operator. We describe interactions as a zero-range pairwise interaction potential $V(x-x') \equiv g \delta(x-x')$ \cite{Olshanii} as in the Lieb-Liniger model, with $g>0$ to describe repulsive interactions. We will assume that the system is homogeneous in the longitudinal direction and adopt periodic boundary conditions.
We focus on the concept of superfluidity as probed by a weak potential barrier stirred in the Bose fluid. A possible experimental setup for this barrier is a Gaussian laser beam whose waist $w$ moves at a velocity $v$ in the frame of the fluid, leading to a perturbing Hamiltonian
\begin{equation}
\label{Pot}
 H_1=\int_0^L \mathrm{d}x \sqrt{\frac{2}{\pi}}\frac{U_b}{w}e^{-\frac{2(x-vt)^2}{w^2}}\psi^\dagger(x)\psi(x),
\end{equation}
Factors have been chosen so that, in the limit $w\to 0$, we recover the results found in \cite{Pitaevskii} for a delta-potential.
We also assume that $U_b$ is low enough so that $H_1$ can be treated perturbatively within the linear response theory.

As in classical physics, the drag force $F$ acting on the fluid is linked to the mean dissipated energy per unit time by the definition
\begin{equation}
 \langle \overline{\dot{E}} \rangle \equiv -F \cdot v.
\end{equation}
In a superfluid, displaying frictionless flow, we expect no energy dissipation and the drag force vanishes in a certain range of velocities. Notice that energy dissipation can be directly probed in experiments \cite{Ketterle, David, Dalibard}.
Using linear-response theory and the fluctuation-dissipation theorem, in the thermodynamic limit, the most general form of the drag force in 1D is \cite{Caux} (see appendix \ref{Flinresp} for details)
\begin{equation}
\label{drag}
F=\frac{1}{2\pi\hbar}\int_0^{+\infty}\mathrm{d}q|U(q)|^2 qS(q,qv)(1-e^{-\beta\hbar qv}),
\end{equation}
where $U(q)$ is the Fourier transform of the barrier potential $U(x)\equiv \sqrt{\frac{2}{\pi}}\frac{U_b}{w}e^{-2\frac{x^2}{w^2}}$ and
\begin{equation}
\label{DSF}
S(q,\omega)=\int_{-\infty}^{+\infty}\!\!\!\int_{-\infty}^{+\infty}\!\!\!\mathrm{d}x \mathrm{d}t e^{i(\omega t-qx)} \langle \delta n(x,t) \delta n(0,0) \rangle_0
\end{equation}
is the dynamic structure factor \cite{Stringari}, giving the weight of the excitation spectrum. $\langle \dots \rangle_0$ indicates the quantum statistical average with respect to the unperturbed Hamiltonian $H_0$, and $\delta n(x,t) \equiv n(x,t)-n_0$ represents the local fluctuations of the density operator. The dynamic structure factor is linked to the Fourier transform of the density-density linear response function by the fluctuation-dissipation theorem.
In Eq.~(\ref{drag}) the dynamic structure factor is integrated along the line $\omega=qv$ in the wavenumber-energy plane: if $S(q,\omega)$ takes arbitrarily small values along this line, i.e. no collective excitations are possible, then the drag force vanishes and the flow is superfluid. We note that this is a sufficient but not necessary condition since there are other factors in the integrand in Eq.~(\ref{drag}).

In the following, we shall determine the drag force acting on the fluid by computing the time-dependent density-density correlation functon, evaluate its Fourier transform with respect to time and space to get the dynamic structure factor (\ref{DSF}), and eventually use Eq.~(\ref{drag}).
We expect the drag force to depend on the stirring velocity $v$, the temperature $T$, the waist $w$ of the barrier and its strength $U_b/w$, the interaction between atoms $g$, and the equilibrium linear density $n_0$.
The strength of boson-boson interactions is expressed through the dimensionless parameter
\begin{equation}
\gamma \equiv \frac{E_{int}}{E_{kin}}=\frac{mg}{n_0\hbar^2},
\end{equation}
ratio of the interaction to kinetic energy of the atomic gas in the equilibrium Hamiltonian $H_0$. It can be fine-tuned experimentally by Feshbach or confinement-induced resonances. We will focus on the strongly interacting regime $\gamma \gg 1$.
In the limit $\gamma\to +\infty$ of infinitely interacting bosons we use an exact solution \cite{Girardeau}. At arbitrary interaction strength we use a low-energy effective Hamiltonian given by the linear Luttinger model.

\section{Limiting case $\Large{\gamma\to +\infty}$: the Tonks-Girardeau gas}
\label{TG}

\subsection{Exact Bose-Fermi mapping}
The limiting case $\gamma \to +\infty$ is known as the Tonks-Girardeau gas \cite{Girardeau}. A well-known peculiarity of one-dimensional systems is that in many respects, hard-core bosons behave like free fermions \cite{Girardeau,Lieb2,Yang}. This phenomenon, related to statistical transmutation, is due to the fact that infinite repulsive interactions impose the many-body wavefunction to vanish wherever there is a contact between two particles, which is the same condition occuring in a Fermi gas owing to antisymmetry of the wavefunction. Namely, one can write the many-body wavefunction of the hard-core Bose gas $\psi^G$, where the superscript $G$ stands for ``Tonks-Girardeau'', in terms of the one of a free Fermi gas denoted by $\psi^F$, as $\psi^G(x_1,\dots,x_N)=\prod_{(i,j)} {\rm sign}(x_i-x_j)\psi^F(x_1,\dots,x_N)$. The Bose-Fermi mapping has been also demonstrated at finite temperature \cite{Yang}, namely, the thermal average of an observable $O$ for the bosonic gas can be obtained as
\begin{equation}
\label{BF}
\langle O \rangle=\frac{1}{Z}\sum_n w_n \langle \psi_n^F|A^{-1}OA|\psi_n^F\rangle,
\end{equation}
where $A=e^{-i\pi\int_{-\infty}^x \mathrm{d}x' n(x')}$ is the mapping operator expressing the Jordan-Wigner transformation from bosons to fermions, $w_n=e^{-\beta E_n}$ are the thermal weights and $Z$ is the partition function. From Eq.~(\ref{BF}) follows that the particle density and density-density correlation functions of a Tonks-Girardeau gas coincide with the fermionic ones, both at zero and finite temperature, since in this case $[O,A]=0$. In the following we shall exploit this property to determine the dynamic structure factor of the TG gas.

\subsection{Dynamic structure factor}
\label{DSFTonks}

We recall first the zero temperature results. Computing fermionic density-density correlations using Wick's theorem yields, after Fourier transform, the well-known result for the dynamic structure factor $S^G$ of the Tonks-Girardeau gas \cite{Vignolo,Caux}
\begin{equation}
\label{TonksDSF}
S^{G}(q,\omega)=\frac{m}{\hbar |q|}\Theta \left[\omega_+(q)-\omega\right]\Theta \left[\omega-\omega_-(q)\right],
\end{equation}
where $\Theta$ is the Heaviside distribution and $\omega_+(q)=\frac{\hbar}{2m}(q^2+2qk_F)$, $\omega_-(q)=\frac{\hbar}{2m}|q^2-2qk_F|$ are the limiting dispersion relations bounding the area where particle-hole excitations can occur, due to energy conservation. Recalling the Bose-Fermi mapping, it is natural to call Fermi wavevector the quantity $k_F=\pi n_0$, with $n_0=N_0/L$.

At finite temperature, the linear density-density response function for the Tonks-Girardeau gas in the Fourier space is given by Lindhard's expression \cite{Ashcroft}
\begin{equation}
\label{LinR}
\chi_{nn}(q,\omega)=\frac{1}{L}\sum_k\frac{n_F(k)-\!n_F(k+q)}{\hbar\omega+\epsilon(k)-\epsilon(k+q)+i0^+},
\end{equation}
where $n_F(k)=\frac{1}{e^{\beta[\epsilon(k)-\mu]}+1}$ is the Fermi-Dirac distribution and $\epsilon(k)=\frac{\hbar^2 k^2}{2m}$ the dispersion relation. Using the fluctuation-dissipation theorem, we deduce from Eq.~(\ref{LinR}) the expression of the finite-temperature dynamic structure factor in the thermodynamic limit:
\begin{eqnarray}
\!\!\!\!\!\!\small{S^{G}_T(q,\omega)=\int_{-\infty}^{+\infty}\!\!\!\!\mathrm{d}k\frac{n_F(k)\!-\!n_F(k+q)}{1-e^{-\beta\hbar\omega}}\delta\left[\omega-\omega_q(k)\right]}
\end{eqnarray}
where $\omega_q(k)\equiv \frac{1}{\hbar}[\epsilon(k+q)-\epsilon(k)]$ and $\delta$ is the Dirac distribution. This is equivalent to
\begin{equation}
\small{S^{G}_T(q,\omega)\!=\!\int_{-\infty}^{+\infty}\! \!\!\mathrm{d}k\ n_F(k)\left[1-n_F(k+q)\right]\delta\left[\omega-\omega_q(k)\right]},
\end{equation}
which is easier to interpret physically in terms of particle-hole excitations. Either of them can be used to obtain
\begin{equation}
\label{DSFTG}
S^{G}_T(q,\omega)=\frac{m}{\hbar|q|} \frac{n_F\left(\frac{\hbar\omega-\epsilon(q)}{\hbar^2 q/m}\right)-n_F\left(\frac{\hbar\omega+\epsilon(q)}{\hbar^2 q/m}\right)}{1-e^{-\beta\hbar\omega}}.
\end{equation}

\begin{figure}
\includegraphics[width=4.2cm, keepaspectratio, angle=0]{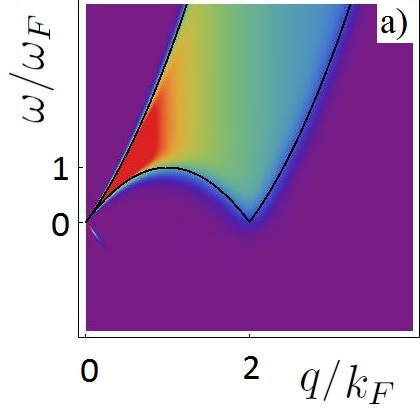}
\includegraphics[width=4.2cm, keepaspectratio, angle=0]{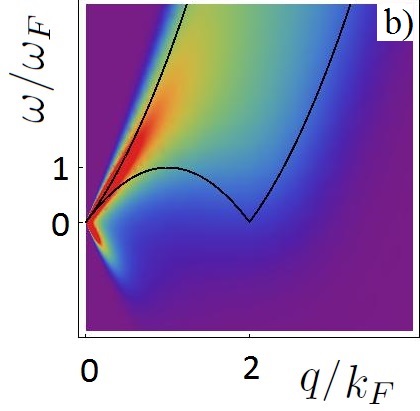}
\includegraphics[width=4.2cm, keepaspectratio, angle=0]{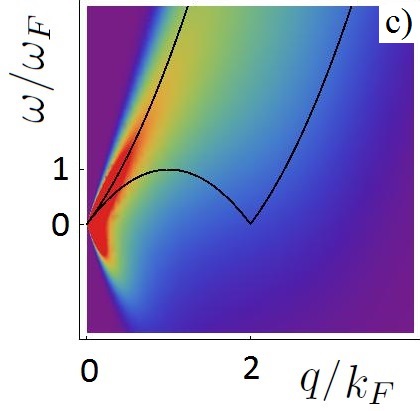}
\includegraphics[width=4.2cm, keepaspectratio, angle=0]{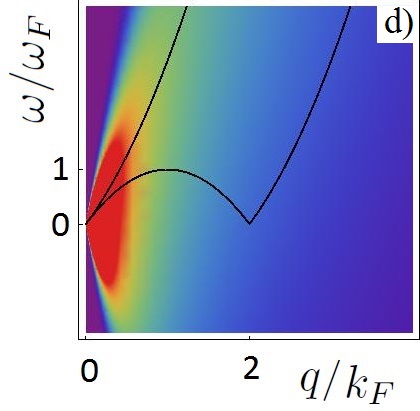}
\caption{(Color online) Dynamic structure factor $S^{G}_T(q,\omega)$ of the Tonks-Girardeau gas in the thermodynamic limit, in units of $S^{
G}(2k_F,0)$, for several dimensionless temperatures $T/T_F=0.1, 0.5, 1$ and $4$ in panels a), b), c) and d) respectively, where $T_F$ is the Fermi temperature. Frequencies $\omega$ are expressed in units of $\omega_F\equiv \epsilon_F/\hbar$, where $\epsilon_F=\hbar^2k_F^2/2m$ is the Fermi energy, and wavenumbers $q$ in units of the Fermi wavevector $k_F$. Black solid lines correspond to the limiting dispersion relations $\omega_+$ and $\omega_-$, in units of $\omega_F$, defining the excitation domain at $T=0$.}
\label{tonks}
\end{figure}

In order to obtain the finite-temperature dynamic structure factor, we have first found the temperature dependence of the chemical potential as detailed in Appendix \ref{muT}.
The results are shown in Fig.~\ref{tonks}. $S(q,\omega)$ is very sensitive to temperature.  Non-vanishing contributions spread beyond the particle-hole excitation spectrum boundaries at finite temperature, since the dynamic structure factor includes thermally-activated excitations. The latter can even occur at $\omega<0$, meaning that collective excitations can be emitted. The quasi-linear shape of the spectrum near the origin and the umklapp point $(2k_F,0)$ at $T=0$ fades out at temperatures larger than $0.1 T_F$, where $T_F \equiv \frac{\hbar^2 k_F^2}{2m k_B}$ is the Fermi temperature. Increasing the temperature contributes to breaking down the imbalance between the $\omega>0$ and $\omega<0$ domains, see panel d) of Fig.~\ref{tonks}.

\subsection{Drag force}
Using the exact expressions for the dynamic structure factor, the next step is to compute the drag force. We start recovering the known result in the case $w=0$, $T=0$.
Combining Eq.~(\ref{drag}) and Eq.~(\ref{DSFTG}), we find, in agreement with \cite{Caux},
\begin{equation}
\label{basic}
F^{G}(v)=\frac{2 U_b^2 n_0 m}{\hbar^2}\left[\Theta(v-v_F)+\frac{v}{v_F}\Theta(v_F-v)\right]\!,
\end{equation}
where $v_F\equiv \frac{\hbar k_F}{m}$ is the Fermi velocity. The drag force is linear with the barrier velocity if $v<v_F$ and saturates if $v>v_F$. As we will see below, this saturation is an artifact due to several theoretical simplifications. Equation~(\ref{basic}) shows that the drag force never vanishes if the velocity of the perturbing potential is finite, meaning that energy dissipation will occur as long as a barrier is driven across the fluid. Hence, the Tonks-Girardeau gas is not superfluid according to the drag force criterion.

We next generalize our calculations to the case of a finite laser waist, still assuming that $T=0$. The Fourier transform of the potential in Eq.~(\ref{Pot}) reads
\begin{equation}
\label{TFPot}
U(q)=U_b e^{-\frac{q^2 w^2}{8}},
\end{equation}
so that the drag force is readily obtained as
\begin{equation}
\!\!F_{w}(v)=\frac{U_b^2}{2\pi\hbar}\int_0^{+\infty}\!\!\mathrm{d}q e^{-\frac{q^2 w^2}{4}} q S(q,qv).
\end{equation}
For the Tonks-Girardeau gas, this yields the analytic expression at $T=0$,
\begin{eqnarray}
\label{FwTG}
\!\!\!\!\!\!\!\!\small{\frac{F_{w}^{G}(v)}{F^{G}(v_F\!)}\!=\!\!\frac{1}{2 w k_F}\!\!\left\{\!h\!\left[w k_F\!\left(\!1\!+\!\frac{v}{v_F}\!\right)\!\right]\!\!-\!h\!\left[w k_F\!\!\left|1\!-\!\frac{v}{v_F}\!\right|\right]\!\right\}}
\end{eqnarray}
where $h(x) \equiv \int_0^x \mathrm{d}u e^{-u^2}$.

\begin{figure}
\includegraphics[width=7cm, keepaspectratio, angle=0]{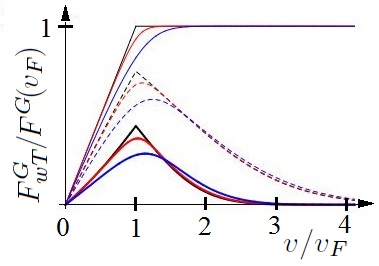}
\caption{(Color online) Drag force $F_{wT}^{G}(v)$ in units of $F^{G}(v_F)$ in a Tonks-Girardeau gas as a function of the dimensionless barrier velocity $v/v_F$. Solid lines stand for a dimensionless waist $w k_F=0$, dashed lines for $w k_F=0.5$ and thick lines for $w k_F=1$. For a given set of curves, temperature increases from $0$ to $0.1 T_F$ to $0.5 T_F$ from top to bottom, in black, red and blue respectively .}
\label{fT}
\end{figure}

We next discuss thermal effects on the drag force. We first treat the limit case of an infinitely thin barrier ($w\to 0$) at finite temperature. The drag force reads
\begin{equation}
\frac{F_T^{G}}{F^{G}(v_F)}\!=\!\frac{1}{2}\sqrt{\frac{T}{T_F}}\int_0^{\beta mv^2/2}\!\!\!\!\frac{d\epsilon}{\sqrt{\epsilon}(e^{\epsilon-\beta\mu(T)}+1)}.
\end{equation}
The integral can easily be evaluated numerically. Notice that, at very low temperatures, it is very close to the Fermi integral $F_{-1/2}(\beta\epsilon_F)$ \cite{Temme,Cloutman}.
The most general case of finite waist and temperature is obtained by inserting Eqs~(\ref{DSFTG}) and (\ref{TFPot}) in Eq.~(\ref{drag}) and then evaluating it numerically.
Figure \ref{fT} shows the drag force as a function of the potential barrier velocity for several values of the temperature and waist. As a main result, thermal effects broaden the curves around the Fermi velocity, and at low velocity the drag force remains linear. If $w=0$, $F(v\ll v_F)\simeq \frac{v}{v_F}\frac{1}{1+e^{-\beta\mu}}$, thus measuring the slope near the origin yields the chemical potential. The finite width of the laser beam should be compared to the interparticle distance: the drag force decreases at increasing $w\pi n_0$. The saturation of the drag force disappears at finite barrier width.

Now we proceed to study the case of finite but large interaction strengths.

\section{Finite interaction strength $\gamma$: The Luttinger liquid model}
\label{LL}

\subsection{The model}

The Luttinger liquid model was introduced to describe interacting fermions in 1D, where the Fermi liquid paradigm breaks down \cite{Tomonaga, Efetov, Larkin, Mattis, Luther, Haldane, Haldane1}. Owing to statistical transmutation, the same approach handles as well one-dimensional interacting bosons \cite{Cazalilla,Giamarchi}. At $T=0$, Luttinger liquids belong to the universality class of 1D systems with gapless, linearly-dispersive excitations. The Lieb-Liniger model belongs to this class at low enough energy, since its dynamic structure factor can be linearized around the origin and the umklapp point in the $(q,\omega)$ plane.
Therefore, we describe the bosonic fluid by the effective Tomonaga-Luttinger Hamiltonian \cite{Cazalilla}
\begin{equation}
\label{HL}
H^{LL}=\frac{\hbar v_s}{2\pi} \int_0^L \mathrm{d}x\left[K(\partial_x\phi)^2+\frac{1}{K}(\partial_x\theta)^2\right],
\end{equation}
where the superscript $LL$ will hereafter denote quantities computed for a Luttinger liquid. In Equation~(\ref{HL}), $\phi(x)$ is the phase field in the phase-density representation of the bosonic field operator $\psi(x) \equiv \sqrt{n(x)}e^{i\phi(x)}$, $\theta(x)$ is a field related to the number of particles between the origin and the position $x$ and whose derivative has a peak whenever a particle is encountered, satisfying the commutation relation $[\partial_x\theta(x),\phi(x')]=i\pi \delta(x-x')$. The isothermal sound velocity $v_s$ and the dimensionless Luttinger parameter $K=\hbar\pi n_0/mv_s$ are taken as an input of the theory. They can be extracted experimentally since they are linked to measurable quantities (compressibility, density), or from a microscopic model whose low-energy limit is the Luttinger liquid we consider.

\begin{figure}
\includegraphics[width=7cm, keepaspectratio, angle=0]{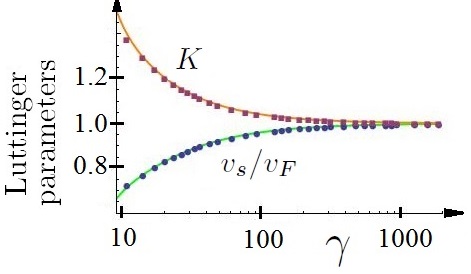}
\caption{(Color online) Luttinger parameters $v_s$ in units of $v_F$ (blue circles) and $K$ (dimensionless, purple squares) as a function of the dimensionless interaction strength $\gamma$ at $T=0$. The solid lines are the asymptotic expansions at large interactions.}
\label{Bethe}
\end{figure}

For 1D bosons described by the Lieb-Liniger Hamiltonian (\ref{LiebLiniger}), we have obtained the Luttinger parameters by numerically solving the Bethe-Ansatz equations (cf. Appendix \ref{BetheAnsatzsolve}) both at zero \cite{Lieb} and finite temperature \cite{Yang}. Results at zero temperature are shown in Fig. \ref{Bethe}, together with the asymptotic expansion at large interactions, $v_s(\gamma)/v_F\simeq 1-\frac{4}{\gamma}+\frac{12}{\gamma^2}+(\pi^2-6)\frac{16}{3\gamma^3}-(2\pi^2-3)\frac{80}{3\gamma^4}$ \cite{Zvonarev}.
The Tonks-Girardeau gas limit $\gamma \to +\infty$ treated in the previous section corresponds to the values $K=1$ and $v_s=v_F$ in the Luttinger liquid description at zero temperature. For bosons with repulsive interactions one has $K>1$, whereas for repulsive fermions $K<1$. The case $K=1$ corresponds to both infinitely-interacting bosons and free fermions, owing to the Bose-Fermi mapping.

\begin{figure}
\includegraphics[width=7cm, keepaspectratio, angle=0]{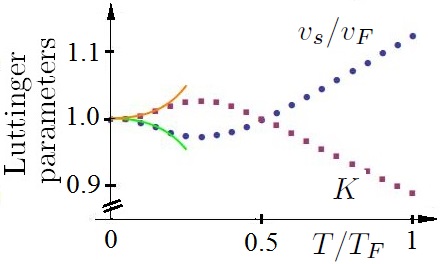}
\caption{(Color online) Temperature dependence of the Luttinger parameters $v_s$ in units of $v_F$ (blue circles) and $K$ (dimensionless, purple squares) in the Tonks-Girardeau regime. Solid lines are low-temperature asymptotical expansions, extracted from the Sommerfeld expansion.}
\label{vs(T)}
\end{figure}

The Luttinger parameters generally depend on temperature. To illustrate this, we compute the temperature-dependence of the sound velocity in the Tonks-Girardeau regime
as extracted from the static structure factor \cite{ChernyCaux} $S(q)\equiv\int_{-\infty}^{+\infty}\mathrm{d}\omega S(q,\omega)$ according to the compressibility sum rule \cite{Nozieres}
\begin{equation}
\label{sumrule}
\lim_{q \to 0} S_T(q)=2k_F\frac{k_BT}{\hbar^2n_0}\left(\frac{\partial n_0}{\partial \mu}\right)_T=2k_F\frac{k_BT}{mv_s^2},
\end{equation}
where $S^G_T(q)$ is evaluated numerically using Eq.~(\ref{DSFTG}) and the temperature dependence of the chemical potential (see again Appendix \ref{muT}). Then we use the relation $Kv_s=\frac{\hbar\pi n_0}{m}$, stemming from Galilean invariance, hence also valid at finite temperature, to obtain the Luttinger parameter $K$. We also extract $v_s$ analytically from the Sommerfeld expansion of the chemical potential at low temperature, yielding for $T\ll T_F$: $\frac{v_s(T)}{v_F} \simeq 1\!-\!\frac{\pi^2}{24}\left(\frac{T}{T_F}\right)^2\!-\!\frac{31\pi^4}{576}\left(\frac{T}{T_F}\right)^4$.
Our results are shown in Fig.~\ref{vs(T)}. Note in particular that for $T \lesssim 0.5T_F$, one has $K>1$. Although the effect is small at low temperatures, as we shall see in Sec.~\ref{DragT} below it is important to include the temperature corrections in the Luttinger parameters in order to find agreement with the Tonks-Girardeau exact solution.

We proceed by diagonalizing the Luttinger Hamiltonian (\ref{HL}), using a mode expansion of $\theta$ and $\phi$ over bosonic fields obtained for periodic boundary conditions \cite{Cazalilla}
\begin{equation}
\theta(x)=\theta_0+\frac{\pi x}{L}(N-N_0)+\sum_{q\neq 0}\!\left|\frac{\pi K}{2qL}\right|^{\frac{1}{2}}e^{-\frac{\epsilon |q|}{2}}(e^{iqx}b_q+h.c.)
\label{modetheta}
\end{equation}
and
\begin{equation}
\label{modephi}
\small{\!\!\!\!\phi(x)\!=\!\phi_0\!+\!\frac{\pi x}{L}J\!+\!\!\sum_{q\neq 0}\!{\rm sign}(q)\!\!\left|\frac{\pi}{2KqL}\right|^{\frac{1}{2}}\!e^{-\frac{\epsilon |q|}{2}}(e^{iqx}b_q\!+\!h.c.)},
\end{equation}
where $h.c.$ means hermitian conjugated, with $b_q$ a bosonic field operator satisfying the commutation relation $[b_q,b^\dagger_{q'}]=\delta_{q,q'}$ and $\epsilon$ is a model-dependent high-momentum cut-off. The zero-mode terms in Eqs~(\ref{modetheta}) and (\ref{modephi}) contain respectively the particle number operator $N$ and the topological number operator $J$, which we shall drop as we do not describe angular-momentum carrying states, as well as their conjugate zero-mode fields $\theta_0$ and $\phi_0$. The latter, though may play a role in finite-size systems \cite{Loss,Cominotti}, do not contribute to the correlation function of interest here. Once diagonalized, the Luttinger Hamiltonian (\ref{HL}) reads
\begin{equation}
 H^{LL}=\sum_{q\neq 0} \hbar \omega(q) b_q^\dagger b_q
\end{equation}
with $\omega(q)=|q|v_s$, i.e. the bosonic field obeys a linear dispersion relation. These phonons are collective oscillations of the phase and density fields and well describe the excitations of the bosonic fluid at low momentum.
We proceed to compute the density-density correlations, the dynamic structure factor and the drag force.

\subsection{Density-density correlations}

We start by recovering the zero temperature result. Using the bosonized expression of the density operator: $n(x)=\frac{1}{\pi}\partial_x\theta(x)\sum_{m=-\infty}^{+\infty}e^{2im(\theta(x)-\pi n_0 x)}$ \cite{Cazalilla} and the mode expansion (\ref{modetheta}) we obtain in agreement with \cite{Cazalilla, Mazzanti}, up to leading orders in the index $m$,
\begin{eqnarray}
\label{nnT0}
&&\frac{\langle \delta n(x,t)\delta n(0,0)\rangle^{LL}}{n_0^2} \simeq \nonumber \\
&&-\: \frac{K}{4 k_F^2}\left[\frac{1}{(x-v_s t+i\epsilon)^2}+\frac{1}{(x+v_s t-i\epsilon)^2}\right] \nonumber \\ 
&&+\: A_1(K)\frac{\cos(2k_F x)}{n_0^{2K}\left[(x-v_s t+i\epsilon)(x+v_s t-i\epsilon)\right]^K},
\end{eqnarray}
where $A_1$ is a non-universal parameter. We can express it as $A_1(K)=2(\epsilon n_0)^{2K}$ where the cut-off $\epsilon$ depends on the interaction strength $\gamma$ and thereby on the Luttinger parameter $K$. As we shall see below, the first term is related to the dynamic structure factor near the origin in the $(q,\omega)$ plane, while the second one has a finite contribution in the vicinity of the umklapp point $(2k_F,0)$. The algebraic decay of the correlations is a signature of quasi-long-range order. We have kept only the two leading terms since larger orders decay with an increasingly large coefficient $Km^2$, with $m$ a non-negative integer. This expression does not depend on the statistics, except through the values of $K$ and $v_s$, which span different ranges for repulsive fermions and bosons.

Using the same approach, we generalize the previous result to finite temperature (details are found in Appendix \ref{finiteT})
\begin{eqnarray}
\label{nnT}
&&\frac{\langle \delta n(x,t)\delta n(0,0)\rangle_T^{LL}}{n_0^2}\simeq \nonumber \\
&&-\: \frac{K}{4 k_F^2}\frac{\pi^2}{L_T^2}\left\{\frac{1}{\sinh^2\left[\frac{\pi(x-v_s t)}{L_T}\right]}+\frac{1}{\sinh^2\left[\frac{\pi(x+v_s t)}{L_T}\right]}\right\} \nonumber \\
&&+\: \frac{2\cos(2k_Fx)}{\left\{\frac{L_T^2}{\pi^2\epsilon^2}\sinh\left[\frac{\pi (x-v_s t)}{L_T}\right]\sinh\left[\frac{\pi (x+v_s t)}{L_T}\right]\right\}^{K}},
\end{eqnarray}
where $L_T\equiv \beta \hbar v_s$ is a thermal length. This expression is valid in the limit $x\pm v_s t, v_s t, L_T \gg \epsilon$, and agrees with \cite{Giamarchi}. Here again we have kept only the two leading terms in the dynamic factor because of the exponential decay of higher-order terms with a $Km^2$ exponent.

\subsection{Dynamic structure factor}

At $T=0$, the Fourier transform of Eq.~(\ref{nnT0}) yields the dominant terms of the dynamic structure factor. Since the latter is symmetric with respect to the $\omega$-axis, we write the result for $q>0$
\begin{eqnarray}
\label{DSFLL}
&&S^{LL}(q,\omega)=K|q|\delta[\omega-\omega(q)] \nonumber \\
&&+B_1(K)\left[\omega^2-(q-2k_F)^2v_s^2\right]^{K-1}\Theta[\omega-|q-2k_F|v_s] \nonumber \\
&&\equiv S_0^{LL}(q,\omega)+S_1^{LL}(q,\omega),
\end{eqnarray}
where $B_1(K)\!\equiv\!\frac{A_1(K)}{(2n_0v_s)^{2\lbrace K-1\rbrace}}\frac{\pi^2}{\Gamma(K)^2}\frac{1}{v_s}$ is a non-universal coefficient. In the dynamic structure factor, $S_0$ displays a sharp feature in correspondence to the linear dispersion $\omega(q)=qv_s$. There are also two linear limiting dispersion relations described by $S_1$, symmetric with respect to the $q=2k_F$ line, forming a triangular shape at the umklapp point $(2k_F,0)$. The slopes of the limiting dispersions in $S_0$ and $S_1$ depend on the interaction strength via the interaction-dependent sound velocity. Hence, measuring them for the Lieb-Liniger gas is a way to determine $v_s$.

\begin{figure}
\includegraphics[width=6cm, keepaspectratio, angle=0]{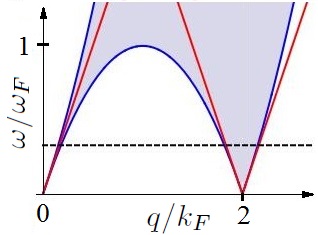}
\caption{(Color online) Definition domain of the dynamic structure factor at $T=0$ in the plane $(q,\omega)$ in units of $(k_F,\omega_F)$. We superimposed the result for a Tonks-Girardeau gas to the result in the Luttinger liquid framework for dimensionless parameters $K=1$, $v_s/v_F=1$. In the latter, the area consists in a line starting from the origin, and the area included in the triangle starting from the umklapp point $(0,2k_F)$. The upper energy limit of potential validity of the Luttinger liquid model is approximately given by the dashed line.}
\label{VS}
\end{figure}

In Fig.~\ref{VS} we plot the definition domain of the dynamic structure factor of a Luttinger liquid with parameters $K=1$ and $v_s=v_F$, superimposed on the result found in the previous section for a Tonks-Girardeau gas (\ref{TonksDSF}). The comparison shows the domain in the $(q,\omega)$ plane where the Luttinger liquid description is valid: namely, at low energy, at small $q$ and for $q\simeq 2k_F$, close to the umklapp point, where the curvature of the dispersions $\omega_{\pm}(q)$ can be neglected. By comparing Eq.~(\ref{DSFLL}) with Eq.~(\ref{TonksDSF}), we notice that $S^{LL}(q,\omega)$ close to the umklapp point does not reproduce the $1/q$ behavior found in the exact model, yet there is quantitative agreement with less than $10\%$ error all the same, provided that $\omega \lesssim 0.33 \omega_F$. This result coincides with the limiting domain found graphically on Fig.~\ref{VS} by looking at the curvature. In this analysis, the cut-off $\epsilon$ is estimated by stating that the matching is optimal on the $q=2k_F$ line, yielding $B_1(K=1)=\frac{1}{2v_F}$ and $\epsilon(K=1)=\frac{1}{2k_F}$ respectively.

We derive next the dynamic structure factor of a Luttinger liquid at finite temperature. The term $S_{0T}^{LL}(q,\omega)$ is more easily computed in one step rather than by evaluating the real-space density-density correlation functions as an intermediate result.
Bosonizing from scratch at finite temperature, we find the first contribution to the dynamic structure factor (details can be found in Appendix \ref{SfiniteT})
\begin{eqnarray}
\label{S0LLT}
\!\!&& S_{0T}^{LL}(q,\omega)= \nonumber \\
\!\!&& \small{\frac{K|q|}{1\!-\!e^{-\beta\hbar\omega(q)}}\!\!\left\lbrace\delta[\omega-\omega(q)]+e^{-\beta\hbar\omega(q)}\delta[\omega+\omega(q)]\right\rbrace}.
\end{eqnarray}
By comparing with the exact results found in Sec.~\ref{DSFTonks} we can now discuss the issue of the temperature range over which a Luttinger liquid may yield results close to those found with the Lieb-Liniger model. Equation~(\ref{S0LLT}) shows that in the Luttinger liquid framework the dynamic structure factor at finite temperature remains linear near the origin, and has two branches. The one corresponding to $\omega<0$ has a lower weight than the $\omega>0$ one and disappears at $T=0$. This behavior is observed in the Tonks-Girardeau limit at low temperature (see again Fig.~\ref{tonks}), yet the linear Luttinger liquid theory does not predict the thermal broadening of the dispersion relation, which becomes more and more relevant at increasing temperature, nor the broadening and the curvature due to non-linearities. We estimate that in the infinitely interacting regime, the Luttinger liquid predictions for the dynamic structure factor around the origin are satisfactory for temperatures lower than approximately $0.15 T_F$. At larger temperatures, it is not relevant to linearize the Tonks-Girardeau dynamic structure factor around the origin because the broadening is too pronounced. Furthermore, combining Eq.~(\ref{S0LLT}) with the linearization at small $q$ of the exact dynamic structure factor, we verify that the temperature dependence of the sound velocity is small at low temperature: for $T\lesssim 0.15 T_F$, we find $0.97 \leq v_s/v_F \leq 1.00$, in agreement with Fig.~\ref{vs(T)}.

The expression of the backscattering term $S_{1T}^{LL}(q,\omega)$ in the dynamic structure factor is obtained from the Fourier transform of the density-density correlations. Using the property \cite{Gradshteyn} $\Gamma(-\frac{yz+xi}{2y})\Gamma(1+z)=(2i)^{z+1}y\Gamma(1+\frac{yz-xi}{2y})\int_0^{+\infty}\!\!\mathrm{d}t e^{-tx}\sin^z(ty), Re(yi)>0, Re(x-yzi)>0$, where $\Gamma$ is the complex Euler Gamma function, after some algebra, we obtain (for more details, please refer to Appendix \ref{SfiniteT})
\begin{eqnarray}
\label{S1LLT}
&&\small{S_{1T}^{LL}(q>0,\omega)=C(n_0,v_s,\epsilon,T) e^{\frac{\beta\hbar\omega}{2}}} \nonumber \\
&&\small{B\left[\frac{K}{2}+i\frac{\beta\hbar}{4\pi}(\omega+\tilde{q}v_s),\frac{K}{2}-i\frac{\beta\hbar}{4\pi}(\omega+\tilde{q}v_s)\right]}\nonumber \\
&&\small{B\left[\frac{K}{2}+i\frac{\beta\hbar}{4\pi}(\omega-\tilde{q}v_s),\frac{K}{2}-i\frac{\beta\hbar}{4\pi}(\omega-\tilde{q}v_s)\right]},
\end{eqnarray}
where $C(n_0,v_s,\epsilon,T)\equiv \left(\frac{L_T}{2\pi\epsilon}\right)^{2\lbrace 1-K\rbrace}\frac{(n_0\epsilon)^2}{2v_s}$, $\tilde{q}\equiv q-2k_F$ and $B[x,y]=\frac{\Gamma[x]\Gamma[y]}{\Gamma[x+y]}$ is the Euler Beta function. The case $K=1$ can be computed separately, leading to
\begin{eqnarray}
\label{S1K1LLT}
&&\!S_{1T}^{LL}(q>0,\omega)|_{K=1}\nonumber\\
&&\!=\!\frac{(k_F\epsilon)^2}{2v_s}\frac{e^{\beta\hbar\omega/2}}{\cosh\left[\frac{L_T}{4v_s}(\omega+\tilde{q}v_s)\right]\cosh\left[\frac{L_T}{4v_s}(\omega-\tilde{q}v_s)\right]}
\end{eqnarray}
in agreement with the general case since $\overline{\Gamma[z]}=\Gamma[\overline{z}]$ and $\left|\Gamma\left[\frac{1}{2}+iy\right]\right|^2=\frac{\pi}{\cosh[\pi y]}$.

\begin{figure}
\label{VST}
\includegraphics[width=4cm, keepaspectratio, angle=0]{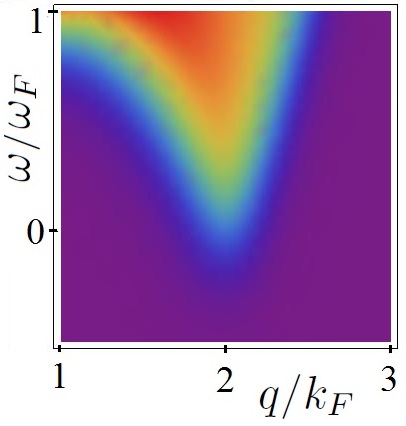}
\includegraphics[width=4cm, keepaspectratio, angle=0]{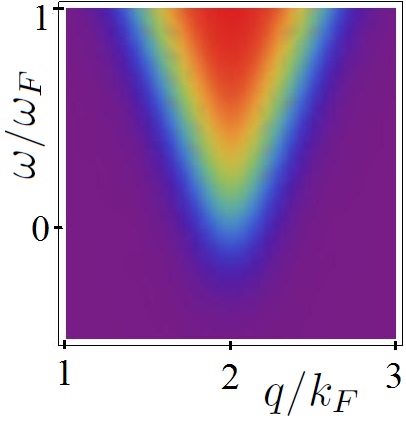}
\caption{(Color online) Dynamical structure factor in units of $S(2k_F,0)_{T=0}$ at $T=0.1T_F$ in the plane $(q,\omega)$ in units of $(k_F,\omega_F)$ in the vicinity of the umklapp point, as predicted for a Tonks-Girardeau gas (left panel) and a Luttinger liquid for dimensionless parameters $K=1.005,v_s/v_F=0.995$ (right panel). The exact temperature-dependence is quite well reproduced by the Luttinger liquid model, differences come mostly from the non-linearities which are not taken into account in the Luttinger liquid framework.}
\end{figure}

As we did for $S_0$, for the backscattering term $S_1$ too we can assess the regime of validity of the Luttinger liquid description at finite temperature by comparing with the Tonks-Girardeau results. Fig.~\ref{VST} shows the dynamic structure factor of a Tonks-Girardeau gas at temperature $T=0.1T_F$ and the one computed for a Luttinger liquid with the appropriate Luttinger parameters at this temperature. The cut-off $\epsilon(0.1T_F)$ is chosen so that the matching is optimal on the $q=2k_F$ line, i.e. they coincide exactly at low energy. Around the umklapp point we find that the Luttinger liquid reproduces quite well the exact thermal broadening, provided that the energy is low enough so that non-linear effects can be neglected, as was already the case at $T=0$.

\subsection{Drag force}
\label{DragT}

Once the dynamic structure factor is known, we can compute the drag force. First, we address the limit $T=0$, $w=0$, which yields
\begin{eqnarray}
\label{FPit}
\!\!&&F^{LL}(v)=\frac{U_b^2}{2\pi\hbar}\int_0^{+\infty}\!\!\mathrm{d}q\!\!\! \quad\!\!qS^{LL}(q,qv) \nonumber \\
\!\!&&=\!\!\frac{U_b^2}{2\pi\hbar}\frac{B_1(K)}{v_s^2}\frac{\sqrt{\pi}\Gamma(K)(2k_Fv_s)^{2K}}{\Gamma(K+1/2)}\frac{\left(v/v_s\right)^{2K-1}}{\left[1\!-\!\left(v/v_s\right)^2\right]^{\!K+1}}
\end{eqnarray}
in agreement with the result found in \cite{Pitaevskii} in the limit $v/v_s \ll 1$: at low velocities, the drag force scales as a power law $v^{2K-1}$.  A comparison with the Tonks-Girardeau result at $K=1$ leads to the determination of the constant $B_1(K=1)=\frac{1}{2v_F}$.

\begin{figure}
\includegraphics[width=7cm, keepaspectratio, angle=0]{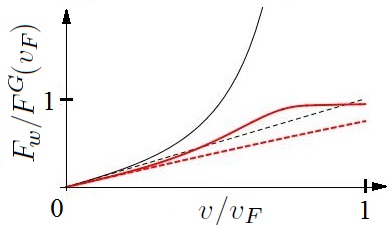}
\caption{(Color online) Drag force in units of $F^{G}(v_F)$ as a function of the velocity $v$ (in units of $v_F$), as predicted for a Tonks-Girardeau gas (dashed lines) and a Luttinger liquid model at dimensionless parameter $K=1$ (solid lines), at $T=0$. Thin black curves correspond to a dimensionless waist $wk_F=0$ and thick red curves to a finite waist $wk_F=0.5$.}
\label{ForceVS}
\end{figure}

Then, we generalize the expression of the drag force to finite laser waist $w$. For the case $K=1$ we find the analytical expression
\begin{equation}
\label{FLLwK1}
F_{w,K=1}^{LL}(v)\!=\!\frac{U_b^2n_0 m}{2\hbar^2k_F^2w^2}\left[e^{\!-\frac{ w^2k_F^2}{(1+v/v_F)^2}}\!-\!e^{\!-\frac{w^2k_F^2}{(1-v/v_F)^2}}\right]\!.
\end{equation}
We use Eqs~(\ref{basic}),(\ref{FwTG}),(\ref{FPit}) and (\ref{FLLwK1}) to plot the curves in Fig.~\ref{ForceVS}, showing that the Luttinger liquid model is able to reproduce exact results in the Tonks-Girardeau regime at low velocities. The drag force is all the better approximated as the potential is wide. It is always linear near the origin, with a slope depending on $w$.

At arbitrary interactions, the expression of the drag force for the case of a finite-width potential is given by
\begin{eqnarray}
\label{FLw}
&&\small{F_{w}^{LL}(v)=\frac{U_b^2}{2\pi\hbar}\frac{B_1(K)}{v_s^2}\frac{\sqrt{\pi}\Gamma(K)(2k_Fv_s)^{2K}}{\Gamma(K+1/2)}\frac{\left(v/v_s\right)^{2K-1}}{\left[1-\left(v/v_s\right)^2\right]^K}} \nonumber \\
&&\small{\frac{1}{w k_F}\!\sum_{k=0}^{+\infty}\!\frac{(\!-1)^k}{k!}\!\left(\!\frac{w k_F}{1\!+\!\frac{v}{v_s}}\!\right)^{2k+1}}\!\!\!\!\!\!\!\!\!\!{_2F_1}\!\left(-1\!-\!2k,\!K;2K;-\frac{2v}{v_s\!-\!v}\!\right)
\end{eqnarray}
where ${_2F_1}$ is the hypergeometric function. We have verified that for $wk_F \lesssim 1$, truncating the sum at low orders yields a very good accuracy.

\begin{figure}
\includegraphics[width=7cm, keepaspectratio, angle=0]{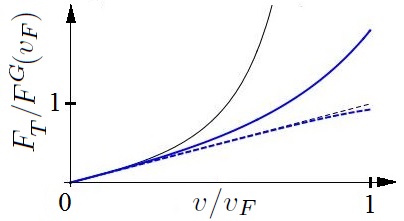}
\caption{(Color online) Drag force in units of $F^{G}(v_F)$ as a function of the velocity $v$ (in units of $v_F$), as predicted for a Tonks-Girardeau gas (dashed lines) and a Luttinger liquid (solid lines), at $w=0$. Thin black curves correspond to $T=0$ and thick blue curves to $T=0.1T_F$.}
\label{ForceVT}
\end{figure}

The effect of temperature on the drag force is obtained by integrating numerically Eq.~(\ref{drag}) with the input of Eqs~(\ref{S0LLT}) and (\ref{S1LLT}). We have plotted in Fig.~\ref{ForceVT} the drag force at $T=0$ and finite temperature as a function of the velocity for a Tonks-Girardeau gas as obtained from the exact solution and the Luttinger liquid approach, where the value of the cut-off $\epsilon$ is chosen by enforcing that the results should coincide at the origin. We note that we find an excellent agreement at lox velocities by taking into account finite-temperature corrections of the LL parameters while the use of zero-temperature values would yield a less precise agreement (not shown) The predictions of the Luttinger liquid and the Tonks-Girardeau gas start differing at the same velocity whether temperature is taken into account or not, essentially due to non-linearities of the Tonks-Girardeau dispersion.

As a main result, we have shown that in the regime of very large interactions, the Luttinger liquid theory is able to reproduce the exact results of the Tonks-Girardeau gas in terms of dynamic structure factor around the umklapp point and drag force at low velocities, even for a potential barrier with a finite width. This allows us to use the Luttinger liquid theory to predict the generic behavior of the drag force at large to intermediate interactions, thus complementing the Bogoliubov approach at weak interactions.

\section{Summary and outlook}

We have used the concepts of dynamic structure factor and drag force to explore theoretically the superfluidity of a system of strongly-interacting bosons in 1D, stirred by a Gaussian laser beam. We have studied the limiting case of infinite interaction strength using the exact Bose-Fermi mapping, and compared it to the predictions of the linear Luttinger liquid model. We have obtained various analytical expressions generalizing known results to finite temperature and finite laser beam width.
The Luttinger liquid model predictions are limited by the non-linearity of the real physical system which are not taken into account within this model. As seen in the exact Tonks-Girardeau solution, at higher energies or at intermediate wavevectors $q\simeq k_F$, beyond-linear Luttinger liquid effects will appear \cite{Khodas, Imambekov, Schmidt, Pirooznia}. These have been experimentally observed \cite{David}. Yet, our work shows that the effects of temperature on the dynamic structure factor around the umklapp point $(q=2k_F, \omega=0)$ and on the drag force are well taken into account. Our results lead to precise estimates for the low-energy behavior of the dynamic structure factor at finite temperature. Generalizations of the drag force to a finite-size potential barrier show that this parameter has a dramatic impact, which is well reproduced by use of the Luttinger-liquid model to treat the Tonks-Girardeau regime. We conclude that, provided that the temperature is low enough and the velocities small, the Luttinger liquid theory
can be used to test the superfluidity of the Lieb-Liniger gas according to the drag force criterion. Moreover, our results are expected to describe a wide range of systems whose low-energy description is a Luttinger Liquid, for instance, the Calogero-Sutherland model \cite{Kawakami, Pustilnik1}.
The barrier potential we have chosen may be realistic to describe future experiments, yet to be even more realistic, finite size effects e.g. in ring traps, should be taken into account, as well as the possible local non-homogeneity of the fluid, e.g. in harmonic traps \cite{Minguzzi}.

\acknowledgments
Frank Hekking acknowledges funding support from the Institut Universitaire de France, Anna Minguzzi from the Handy-Q ERC grant no 25860 and from the ANR project MathostaqANR-13-JS01-0005-0.

%%%%%%%%%%%%%%%%%%%%%%%%%%%%%%%%%%%%%%%%%%%%%%%%%%%%%%%%%%%%%%%%%%%%%%%%%%%%%%%%%%%%%%%%%%%%%%%%%%%%%%%%%%%%%%%%%%%%%%%
\appendix
%%%%%%%%%%%%%%%%%%%%%%%%%%%%%%%%%%%%%%%%%%%%%%%%%%%%%%%%%%%%%%%%%%%%%%%%%%%%%%%%%%%%%%%%%%%%%%%%%%%%%%%%%%%%%%%%%%%%%%%

\section{Drag force in the linear response theory}
\label{Flinresp}

In quantum physics, if a system is described by a Hamiltonian $H=H_0+H_{pert}\equiv H_0+\int d^drA(r)U(r,t)$, where $A$ is a linear operator and $U$ a weak local perturbation, then the average of any observable $B$ coupled to $A$ in the presence of the perturbing potential reads
\begin{equation}
\small{\!\!\!\!\langle B(r,t)\rangle_U\!=\!\langle B\rangle_0\!-\!\!\int_{-\infty}^{+\infty}\!\!\!\!\mathrm{d}t'\!\!\int\!\!d^dr' U(r',t')\chi_{BA}(r\!-\!r',t\!-\!t')}
\end{equation}
to first order, defining the $B-A$ linear response function, denoted by $\chi_{BA}$. Using the interaction representation (denoted by a subscript $I$) and Liouville equation for the density matrix, one can show that if the system without perturbation is Galilean-invariant, then
\begin{equation}
\chi_{BA}(r\!-\!r',t\!-\!t')\!=\!\frac{i}{\hbar}\Theta(t-t')\left\langle[B_I(r\!-\!r',t\!-\!t'),A]\right\rangle_0
\end{equation}
Let $A=B=n$ the density operator. The ensemble average dissipated energy per unit time reads:
\begin{equation}
 \langle\dot{E}\rangle=\left\langle\frac{d}{dt}\int d^d r U(r,t) n(r,t)\right\rangle_U.
\end{equation}
Using the continuity equation and the definition of the density-density response function, with the convention for the Fourier transform $f(q,\omega)\equiv \int_{-\infty}^{+\infty}\int_{-\infty}^{+\infty} dx dt e^{i(\omega t-qx)}f(x,t)$, we obtain:
\begin{align}
\langle\dot{E}\rangle=\int\mathrm{d}^dr \int \frac{\mathrm{d}\omega}{2\pi}i\omega e^{-i\omega t}\int \frac{\mathrm{d}^dq}{(2\pi)^d}e^{iqr}U(q,\omega) \nonumber\\
\int\frac{\mathrm{d}\omega'}{2\pi}e^{-i\omega't}\int\frac{\mathrm{d}^dq'}{(2\pi)^d}e^{iq'r}\chi_{nn}(q',\omega')U(q',\omega').
\end{align}
We then compute the time average, use the property $U(-q,-\omega)=U^{*}(q,\omega)$ and decompose the Fourier transform of the response function into its real and imaginary parts: $\chi_{nn}(q,\omega)\equiv \chi_{nn}'(q,\omega)+i\chi_{nn}''(q,\omega)$. The real part is an even function of its arguments while the imaginary part is odd. Together with the fluctuation-dissipation theorem, this yields:
\begin{equation}
\small{\langle\overline{\dot{E}}\rangle\!=\!-\!\!\int_0^{+\infty}\!\!\frac{\mathrm{d}\omega}{\pi}\!\!\int\!\!\frac{d^dq}{(2\pi)^d}|U(q,\omega)|^2S(q,\omega)\frac{\omega}{2\hbar}(1\!-\!e^{-\!\beta\hbar\omega})}.
\end{equation}
One then specializes to the 1D case and uses the fact that if the perturbing potential is a function of $x\!-\!vt$, then its Fourier transform contains $\delta(\omega-qv)$ due to energy conservation, to find, eventually, Eq.~(\ref{drag}) with $U(q,\omega)\equiv 2\pi U(q)\delta(\omega-qv)$.

\section{Chemical potential of the Tonks-Girardeau gas at finite temperature}
\label{muT}

\begin{figure}
\includegraphics[width=7cm, keepaspectratio, angle=0]{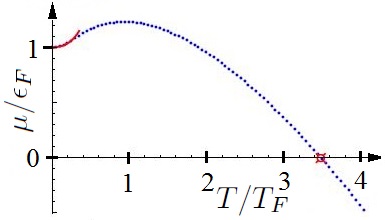}
\caption{(Color online) Chemical potential $\mu$ of the Tonks-Girardeau gas in units of the Fermi energy $\epsilon_F$ as a function of the reduced temperature $T/T_F$, where $T_F=\hbar^2k_F^2/(2mk_B)$ and $k_F=\pi n_0$. The red solid line stands for Sommerfeld's low-temperature expansion to order 4: $\mu/\epsilon_F=1+\frac{\pi^2}{12}(T/T_F)^2+\frac{7\pi^4}{192}(T/T_F)^4$, blue points are the numerical result. The position $(T_0/T_F,0)$ of the isolated point was computed analytically, stating that $\mu$ should vanish at this temperature, yielding $T_0/T_F=4/\pi[(\sqrt{2}-1)\zeta(1/2)]^2\simeq 3.48$, where $\zeta$ is the Riemann zeta function.}
\label{tonksmuT}
\end{figure}

In Fig.~\ref{tonksmuT} we show the temperature dependence of the chemical potential of the Tonks-Girardeau gas, which coincides with the one of a one-dimensional ideal Fermi gas. We note in particular that at low temperature, due to the reduced dimensionality, the chemical potential increases with temperature, at difference from the three-dimensional case.

\section{Solving the Bethe Ansatz equations of the Lieb-Liniger model}
\label{BetheAnsatzsolve}

\begin{figure}
\includegraphics[width=7cm, keepaspectratio, angle=0]{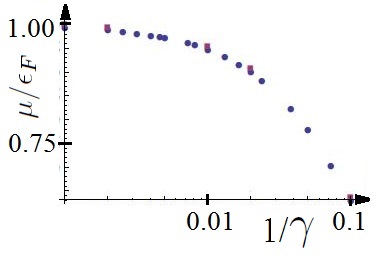}
\caption{(Color online) Chemical potential $\mu$ in units of $\epsilon_F$ as a function of the inverse dimensionless interaction strength $1/\gamma$, at $T=0$ (blue circles) and $T=0.1T_F$ (red squares). }
\label{mu(gammainv)}
\end{figure}

\begin{figure}
\includegraphics[width=7cm, keepaspectratio, angle=0]{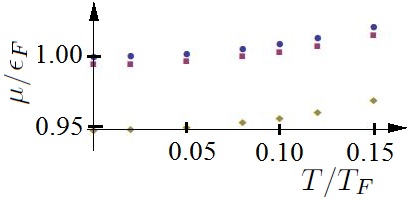}
\caption{(Color online) Chemical potential $\mu$ in units of $\epsilon_F$ as a function of the dimensionless temperature $T/T_F$ for a Tonks-Girardeau gas (blue circles), for a dimensionless interaction strength $\gamma=1000$ (red squares) and $\gamma=100$ (yellow diamonds) respectively. }
\label{mu(t)}
\end{figure}

In \cite{Yang} the Lieb-Liniger Hamiltonian for $N$ bosons is written as
\begin{equation}
H=-\sum_{i=1}^N\frac{\partial^2}{\partial x_i^2}+2c\sum_{i>j}\delta(x_i-x_j).
\end{equation}
The Bethe Ansatz solution for the equation of state at finite temperature in the thermodynamic limit is obtained by solving a system of three coupled equations:
\begin{eqnarray}
\label{fredholm}
&&\epsilon(k)=-\mu(c,T)+k^2\nonumber\\
&&-\frac{T c}{\pi}\!\!\int_{-\infty}^{+\infty}\!\!\frac{\mathrm{d}q}{c^2+(k-q)^2}\ln\left\{1+\exp\left[-\frac{\epsilon(q)}{T}\right]\right\}\!,
\end{eqnarray}
\begin{equation}
\label{eqepsilon}
\small{2\pi\rho(k)\!\left\{1+\exp\left[\frac{\epsilon(k)}{T}\right]\right\}\!=\!1\!+\!2c\!\int_{-\infty}^{+\infty}\!\!\!\!\mathrm{d}q\frac{\rho(q)}{c^2+(k-q)^2}},
\end{equation}
and
\begin{equation}
\label{eqstate}
\int_{-\infty}^{+\infty}\mathrm{d}k\ \rho(k)=n_0,
\end{equation}
where the $k$ are quasi-momenta, $\rho$ is the quasi-momentum distribution function and $\exp[\epsilon(k)/T]\equiv \rho_h/\rho$ with $\rho_h$ the quasi-hole distribution function. With the correspondence $c\equiv \gamma n_0$, we extract $\mu\left(\gamma,T/T_F\right)/\epsilon_F$ from this set of equations.
To find the chemical potential at a given reduced temperature and interaction strength, we proceed as follows. We guess the value of $\mu$ and self-consistently solve Eq.~(\ref{fredholm}) by iteration to find the function $\epsilon(q)$. Once $\epsilon$ is found with enough accuracy, we inject it in Eq.~(\ref{eqepsilon}) and find $\rho$ by iteration. When fluctuations become low enough to be neglected, we check if Eq.~(\ref{eqstate}) is verified up to the chosen accuracy. Most probably, the initial guess for $\mu$ is not precise enough, so this procedure is repeated as many times as needed within a simplex algorithm to find $\mu$ with the chosen accuracy. A few steps are enough to find $\mu/\epsilon_F$ with less than $0.5\%$ error. Our results are shown on Fig.~(\ref{mu(gammainv)}) and Fig.~(\ref{mu(t)}). The chemical potential depends both on the temperature and the interaction strength. When the latter is decreased, the chemical potential increases, while it increases with the reduced temperature. 

\section{Details of the calculation of the density-density correlations in the Luttinger liquid framework at finite temperature}
\label{finiteT}

Here we sketch the derivation of Eq.~(\ref{nnT}). Using $n(x)=\frac{1}{\pi}\partial_x\theta(x)\sum_{m=-\infty}^{+\infty}e^{2im[\theta(x)+k_F x]}$ we have
\begin{eqnarray}
&&\langle n(x,t)n(0,0)\rangle = \frac{1}{\pi^2}\langle\partial_x\theta(x,t)\partial_x\theta(0,0)\rangle \nonumber \\
&&+\frac{1}{\pi^2}\!\!\!\!\!\!\sum_{m,m'=-1,\neq (0,0)}^{1}\!\!\!\!\!\!\!e^{2imk_F x}\langle e^{2im\theta(x,t)}e^{2im'\theta(0,0)}\rangle\!+\!\dots
\end{eqnarray}
which will lead to the zero-order and first-order terms respectively. To find the first-order term we introduce a generating function: $G_{m,m'}(x,t;0,0)\equiv e^{2im\theta(x,t)}e^{2im'\theta(0,0)}$. 
To evaluate $ \langle G_{m,m'}(x,t;0,0) \rangle $ we use the identity $e^{A+B}=e^Ae^Be^{-\frac{1}{2}[A,B]}$, valid for any pair of operators $A$ and $B$ commuting with their commutator, and the property $\langle e^{A} \rangle=e^{\frac{1}{2}\langle A^2 \rangle}$, valid for any linear operator $A$, together with the mode expansion of the field $\theta$ (\ref{modetheta}) and the bosonic commutation relations.
After some algebra, we find
\begin{eqnarray}
&&\langle G_{m,m'}(x,t;0,0) \rangle=e^{2i(m+m')\theta_0} \nonumber \\
&&e^{-\sum_{q\neq 0}\left|\frac{\pi K}{qL}\right|[(m+m')^2+2mm'\{ e^{i[qx-\omega(q)t)]}-1 \} ]}\nonumber \\
&&e^{-\sum_{q\neq 0}\left|\frac{\pi K}{qL}\right|[(m+m')^2+2mm'\{ \cos[qx-\omega(q)t]-1 \} ]n_B(q)},
\end{eqnarray}
where $n_B(q)=\frac{1}{e^{\beta\hbar\omega(q)}-1}$ is the Bose-Einstein distribution for the phonons.
In the thermodynamic limit the non-vanishing contributions are those where $m=-m'$. We find

\begin{eqnarray}
&&\langle G_{m,m'}(x,t;0,0) \rangle=\delta_{m,-m'} \nonumber\\
&&e^{-2K m^2 \int_{q=0}^{+\infty}\frac{\mathrm{d}q}{q}e^{-\epsilon q}\{1-e^{-iqv_s t}\cos(qx)+2[1-\cos(qx)\cos(qv_s t)]n_B(q)\}} \nonumber \\
&&\equiv\delta_{m,-m'}e^{-2K m^2 F(x,t)}.
\end{eqnarray}

To compute $F(x,t)$, one can rewrite $\frac{1}{q}=\int_{0}^{Y \to +\infty}dye^{-qy}$, yielding:
\begin{eqnarray}
&&e^{2F(x,t)}=\frac{(x+v_s t-i\epsilon)(x-v_s t+i\epsilon)}{\epsilon^2}\nonumber \\
&&\prod_{n=1}^{+\infty}\left[1+\frac{(x+v_s t)^2}{(\epsilon+n L_T)^2}\right]\prod_{m=1}^{+\infty}\left[1+\frac{(x-v_s t)^2}{(\epsilon+m L_T)^2}\right].
\end{eqnarray}
The property $\left|\frac{\Gamma(x)}{\Gamma(x-iy)}\right|^2=\prod_{k=0}^{+\infty}\left[1+\frac{y^2}{(x+k)^2}\right]$ yields:
\begin{eqnarray}
&&e^{2F(x,t)}=\frac{x^2-(v_st-i\epsilon)^2}{\epsilon^2}\frac{1}{1+\frac{(x+v_st)^2}{\epsilon^2}}\frac{1}{1+\frac{(x-v_st)^2}{\epsilon^2}} \nonumber \\
&&\left|\frac{\Gamma\left[\frac{\epsilon}{L_T}\right]}{\Gamma\left[\frac{\epsilon}{L_T}\left(1-i\frac{x+v_st}{\epsilon}\right)\right]}\right|^2\left|\frac{\Gamma\left[\frac{\epsilon}{L_T}\right]}{\Gamma\left[\frac{\epsilon}{L_T}\left(1-i\frac{x-v_st}{\epsilon}\right)\right]}\right|^2\!\!,
\end{eqnarray}
then in the limit $v_st,x\pm v_st,L_T \gg \epsilon$, the properties $\Gamma(x)\simeq_{x\to 0}\frac{1}{x}$ and $|\Gamma(iy)|=\frac{\pi}{y \sinh(\pi y)}$ yield Eq.~(\ref{nnT}) after a few rearrangements.

\section{Details of the calculation of the dynamic structure factor in the Luttinger liquid framework at finite temperature}
\label{SfiniteT}

Here we sketch a derivation of Eqs.~(\ref{S0LLT}) and (\ref{S1LLT}).
To prove Eq.~(\ref{S0LLT}), we split $S_{0T}^{LL}(q,\omega)$ into two contibutions:
\begin{eqnarray}
S_{0T}^{LL}(q,\omega)\equiv S_0^{LL,T=0}(q,\omega)+S_0^{LL,T}(q,\omega)
\end{eqnarray}
where the first part is the result at $T=0$ and the second is a purely thermal part. We focus on $S_0^{LL,T}(q,\omega)$, which is more easily computed starting from an intermediate result in the calculation of the density-density correlations. In order to compute
\begin{eqnarray}
&&S_0^{LL,T}(q,\omega)=\frac{K}{4\pi^2}\int_{-\infty}^{+\infty}\int_{-\infty}^{+\infty}\mathrm{d}x\mathrm{d}te^{i(\omega t-qx)}\nonumber\\
&&\int_{q\neq 0}\mathrm{d}q|q|n_B(q)\left(e^{i[qx-\omega(q)t]}+e^{-i[qx-\omega(q)t]}\right)
\end{eqnarray}
we perform the change of variables $u=x-v_s t$ and $v=x+v_s t$. After some algebra we find
\begin{eqnarray}
\!\!\!\!\!\!S_0^{LL,T}(q,\omega)\!=\!\frac{K|q|}{e^{\beta\hbar\omega(q)}\!-\!1}\!\left\{\delta[\omega-\omega(q)]\!+\!\delta[\omega+\omega(q)]\right\}
\end{eqnarray}
thus
\begin{eqnarray}
&&S_{0T}^{LL}(q,\omega)=K|q|\delta[\omega-\omega(q)]\nonumber\\
&&+\frac{K|q|}{e^{\beta\hbar\omega(q)}-1}\left\{\delta[\omega-\omega(q)]+\delta[\omega+\omega(q)]\right\},
\end{eqnarray}
yielding Eq.~(\ref{S0LLT}).
To prove Eq.~(\ref{S1LLT}), we start from the second contribution in Eq.~(\ref{nnT}) and compute its Fourier transform, yielding
\begin{eqnarray}
S_{1T}^{LL}(q,\omega)\propto I_1(a)I_1(b),
\end{eqnarray}

where $I_1(x)\equiv \int_{-\infty}^{+\infty}\mathrm{d}ue^{-ixu}\sinh(u)^{-K}$, $a\equiv \frac{\beta \hbar}{2\pi}[\omega+(q-2k_Fv_s)]$ and $b\equiv \frac{\beta \hbar}{2\pi}[\omega-(q-2k_Fv_s)]$.
Using $\int_0^{+\infty}\mathrm{d}ue^{-ixu}\sinh(u)^{-K}=2^{K-1}\Gamma(\frac{K+ix}{2})\Gamma(1-K)/\Gamma(1+\frac{ix-K}{2})$ and treating the branchcut carefully, we find
\begin{eqnarray}
S_{1T}^{LL}(q\!>\!0,\omega)\!&=&\!\left(\frac{\beta\hbar v_s}{\pi}\right)^{2(1-K)}\!\!\!\!\!\!\epsilon^{2K}\frac{n_0^2}{2v_s}2^{2(K-1)}\Gamma(1-K)^2 \nonumber\\
&\times&\!\!\left[\frac{\Gamma\left(\frac{K+ia}{2}\right)}{\Gamma\left(1\!-\!\frac{K-ia}{2}\right)}\!+\!e^{-iK\pi}\frac{\Gamma\left(\frac{K-ia}{2}\right)}{\Gamma\left(1\!-\!\frac{K+ia}{2}\right)}\right]\nonumber\\
&\times&\!\!\left[\frac{\Gamma\left(\frac{K-ib}{2}\right)}{\Gamma\left(1\!-\!\frac{K+ib}{2}\right)}\!+\!e^{iK\pi}\frac{\Gamma\left(\frac{K+ib}{2}\right)}{\Gamma\left(1\!-\!\frac{K-ib}{2}\right)}\right]\!,
\end{eqnarray}
then after some algebra, using twice the property $\Gamma(z)\Gamma(1-z)=\frac{\pi}{\sin(\pi z)}$, we finally obtain Eq.~(\ref{S1LLT}).
In the case $K=1$, the property $\int_{-\infty}^{+\infty}\mathrm{d}x\frac{e^{-\mu x}}{1-e^{-x}}=\pi[i+cotan(\pi\mu)],0< Re(\mu)<1$ yields Eq.~(\ref{S1K1LLT}) in a few lines of algebra.

\section{Details of the calculation of the drag force in the Luttinger liquid framework at finite barrier width}

Here we derive Eq.~(\ref{FLw}). We need to evaluate: $\int_{q_-}^{q_+} \mathrm{d}q \quad \!\!\!\! q(q-q_-)^{K-1}(q_+-q)^{K-1}\exp\left(-\frac{q^2\rm w^2}{4}\right)$. We split it into two parts using: $q=q\!-\!q_-\!+\!q_-$, then expand the exponential as a power series: $\exp(x)=\sum_{k=0}^{+\infty}\frac{x^k}{k!}$ and expand once again according to $q^{2k}=(q\!-\!q_-\!+\!q_-)^{2k}=\sum_{m=0}^{2k}\binom{2k}{m}(q-q_-)^mq_-^{2k-m}$. To eliminate the remaining integrals, we use the property
$\int_a^b \mathrm{d}x (x\!-\!a)^{\mu-1}(b\!-\!x)^{\nu-1}=(b\!-\!a)^{\mu+\nu-1}B(\mu,\nu)$, where $B$ is the Euler Beta function, then after a resummation and using Pascal's triangle, we are left to evaluate
$W\equiv \sum_{m=0}^{2k+1}\left(\frac{2v/v_s}{1-v/v_s}\right)^m B(K\!+m,K)$, which can be interpreted as the sum of a series with all its terms equal to $0$ from rank $2k+2$ on. We express it in terms of the hypergeometric series ${_2}F_1$, which converges since $\!-\!2k\!-\!1$ is a negative integer:
\begin{equation}
\small{\!\!W\!=\!\frac{2^{1\!-\!2K}\sqrt{\pi}\Gamma(K)}{\Gamma(K+1/2)}{_2}F_1\!\left(\!-\!1\!-\!2k,K;2K;\frac{-2v}{v_s\!-\!v}\right)\!}.
\end{equation}
This readily yields Eq.~(\ref{FLw}).

%%%%%%%%%%%%%%%%%%%%%%%%%%%%%%%%%%%%%%%%%%%%%%%%%%%%%%%%%%%%%%%%%%%%%%%%%%%%%%%%%%%%%%%%%%%%%%%%%%%%%%%%%%%%%%%%%%%%%%%

\end{document}